# SafeStudent Driving: A Multimodal Driver-Safety System to Support Teen Drivers Using Computer Vision and Mobile Sensing

Max Liu[1], Garret Washburn[2]

[1]William P Clements High School, 4200 Elkins Rd, Sugar Land, TX 77479
MaxLiu2k@gmail.com
[2]California Baptist University, 8432 Magnolia Ave, Riverside, CA 92504
softcom.lab.cpp@gmail.com

***ABSTRACT***

*Teen drivers face disproportionately high crash rates, often due to inexperience and inconsistent attention to basic traffic rules. SafeStudent Driving addresses this problem with a multimodal coaching system deployed on both a Raspberry Pi device and a Flutter-based mobile app [1]. The system uses three YOLO-based computer-vision models to detect traffic lights, light-bulb colors, and road signs, an OCR module to read speed-limit values, and an audio model plus IMU data to infer whether turn signals are used during turns [2]. An analysis layer smooths detections over time and triggers prioritized voice prompts through text-to-speech or pre-recorded audio.*

*Key challenges included achieving sufficient model accuracy in varied lighting, running inference fast enough on limited hardware, and designing prompts that inform without distracting the driver [3]. Experiments on sign detection and turn-signal recognition highlight strengths and failure modes, guiding future improvements. Overall, the project demonstrates a practical, low-cost way to help novice drivers build safer habits in real traffic.*

## 1. INTRODUCTION

Learning to drive is one of the riskiest transitions in a teenager's life. Around the world, road traffic crashes kill an estimated 1.19 million people every year and are the leading cause of death for children and young adults ages 5–29 [17]. In the United States, teen drivers ages 16–19 have a fatal crash rate nearly three to four times higher than that of older drivers when adjusted for miles driven [18][19]. Recent national data indicate that this problem is worsening: a large-scale study by Bumper reports that road fatalities involving teen drivers have increased by approximately 25% over the past decade [20].

For novice drivers, the problem is not just a lack of skill, but also inexperience with real-world road conditions, difficulty managing distractions, and limited awareness of how quickly situations can change at an intersection or on a highway. Prior work in smartphone sensing and telematics has shown that sensors such as GPS and accelerometers can be used to detect harsh braking, speeding, and other risky behaviors, offering a way to monitor and coach drivers (Datamotion, 2022) [15]. However, many existing systems either focus on post-incident analysis (such as dashcam footage) or are limited to simple metrics like speed and acceleration without understanding the visual context on the road. Also, many teen drivers drive older and/or cheaper cars that have less advanced safety systems than more modern cars, further contributing to this issue.

This gap matters because the consequences of unsafe driving accumulate over time. A teen who repeatedly rolls through stop signs, misses changing lights, speeding, or forgets turn signals is not just at risk in that moment; those habits can become ingrained and increase the likelihood of serious crashes over years of driving. The long-term impact extends to families, schools, and communities, making it critical to develop tools that actively support safe driving habits from the very beginning of a driver's experience (Papatheocharous, 2023) [9].

Methodology A. Ferreira Júnior et al. used smartphone sensors and machine learning to profile driving style (aggressive, normal, calm) from patterns in acceleration and motion [7]. This is effective for overall behavior scoring but does not explicitly recognize traffic controls or give real-time coaching. My project improves on this by adding camera-based recognition of lights and signs and audio-based turn-signal checks, then turning those into immediate voice prompts for novice drivers.

Methodology B. Brahim et al. proposed a sensor-fusion framework that classifies driving behavior using multiple smartphone sensors [8]. It focuses on building accurate classifiers, but it treats the road environment as a black box and does not target teen drivers specifically. SafeStudent Driving adds explicit perception of stoplights, stop signs, and speed limits, integrating that context into its feedback.

Methodology C. Ramachandra et al. built a Raspberry Pi–based ADAS prototype with lane and collision warnings. It shows Pi hardware is viable but focuses mainly on avoiding crashes. My system extends this idea toward education, emphasizing traffic-control compliance and signaling habits through repeated coaching, not just last-second warnings.

SafeStudent Driving proposes a multimodal driver-assistance system for novice drivers: a camera-based device (implemented as both a Flutter mobile app and a Raspberry Pi 5 unit) that provides real-time audio feedback to encourage safer driving behavior. The system combines a forward-facing camera, accelerometer/IMU, GPS, and microphone to continuously observe both the driving environment and the driver's actions. Using computer-vision models, the camera stream is analyzed to detect traffic lights and regulatory signs (for example, stop signs, yield signs, and speed limit signs). When a speed limit sign is detected, an OCR stage extracts the numeric value so the system can compare the posted limit against the vehicle's estimated speed. In parallel, accelerometer and GPS signals characterize motion patterns associated with risky behavior (such as harsh braking or rapid acceleration), while the microphone listens for turn-signal clicking patterns to help infer whether the driver used a blinker during a turn. These signals are fused by a decision layer that prioritizes and rate-limits prompts, so feedback is timely but not overwhelming. The resulting output is short, actionable spoken guidance such as "You are above the detected speed limit," "Red light detected. Please slow down," or "Do not forget to use your turn signal when turning!" Compared to passive dashcams and post-incident review tools, this approach is proactive and instructional by explicitly recognizing traffic controls in the roadway scene and alerting the driver while recording the reaction for future reference.

In Experiment A, we tested a key blind spot: how reliably SafeStudent Driving detects and interprets regulatory and speed-limit signs in real driving footage, since incorrect readings could produce unsafe feedback. We recorded road clips in three lighting conditions (daytime, sunset, nighttime), manually labeled each clip with ground-truth sign classes and speed-limit values, and then ran the full pipeline (YOLO sign detection followed by speed-limit OCR) on frames sampled at fixed intervals. The most significant finding was that end-to-end performance was strongest in daylight and dropped noticeably at night, largely because OCR accuracy degraded under glare, motion blur, and lower sign visibility at distance.

In Experiment B, we evaluated the turn-signal audio detector's reliability across different cars and noise environments. We recorded sessions per car under silence, music, conversation, and music+conversation, aligned the model's "blinker on/off" outputs with labeled timelines, and computed F1 by condition. Results were highest in silence and declined as cabin noise increased, driven by masking effects and click-timbre differences between vehicles.

## 2. Challenges

In order to build the project, a few challenges have been identified as follows.

### 2.1. Balancing Accuracy and Alert Reliability

A major challenge is ensuring the perception models are accurate enough to help rather than distract the driver. If the system misses stop signs or misread speed limits, users may stop trusting it; if it produces frequent false alerts, the feedback can become annoying and potentially unsafe. To address this, the project can combine locally recorded driving footage with curated public datasets to increase coverage of lighting, weather, distances, and regional sign variations. Accuracy can be improved through structured evaluation on diverse test sets with augmentations, threshold tuning, and confirmation logic that requires consistent detections across multiple frames before triggering an alert. For turn-signal detection, audio classification can be cross-checked with IMU-based turning estimates, so prompts are issued only when multiple signals agree, reducing false positives.

### 2.2. Real-Time Performance on Limited Hardware

Another challenge is meeting real-time performance constraints on limited hardware. Running deep-learning inference on every video frame while also monitoring sensors can overload a Raspberry Pi or smartphone, causing latency that reduces the usefulness of feedback. To mitigate this, the system can separate expensive inference workloads from the user interface by using worker threads or asynchronous processing, ensuring the UI remains responsive. Video processing can be rate-limited by sampling frames at a fixed interval, while audio can be analyzed in short windows to preserve sensitivity to turn-signal clicks. Models can also be optimized for deployment by converting them to efficient formats (for example, TorchScript or TensorFlow Lite) and by using resolution and confidence-threshold settings that balance accuracy with throughput.

### 2.3. Clear and Non-Distracting Driver Alerts

Even if detection is accurate and fast, the system must communicate with the driver in a way that is clear, calm, and minimally distracting. If voice prompts are too frequent, too urgent, or poorly timed, they can increase cognitive load rather than provide support. To address this, the system can use a prioritized voice queue that prevents overlapping messages and enforces a cooldown so repeated detections do not spam the driver. Prompts can be phrased descriptively (for example, "Red light detected ahead") rather than using alarming commands, and alerts can be reserved for confirmed events to avoid unnecessary interruptions. On embedded hardware, speech can be generated with a lightweight text-to-speech engine, while the mobile app can use pre-recorded clips for consistent timing and low latency.

## 3. SOLUTION

SafeStudent Driving links three major components: (1) perception models, (2) decision and fusion logic, and (3) the target hardware platform (Raspberry Pi or mobile).

At startup, the system initializes the camera and sensor inputs (IMU and GPS, plus microphone) and loads the required models for visual recognition and audio classification.

During operation, the camera feed is sampled at a controlled rate to manage compute cost. Each sampled frame is processed to detect relevant traffic controls, including traffic lights and regulatory signage. When a speed-limit sign is found, an OCR step extracts the numeric limit so that the system can interpret the sign in a usable form. In parallel, sensor streams estimate motion patterns and turning behavior, while the microphone analysis determines whether a turn-signal click pattern is present. The fusion layer combines these outputs, smooths noisy detections over short time windows, and applies rules to decide when feedback is warranted (for example, newly confirmed red-light detection, exceeding a posted speed limit, or turning without a blinker). When a rule triggers, the feedback module enqueues a short-spoken prompt and ensures prompts are ordered and non-overlapping. On mobile, the Flutter implementation mirrors the same perception-and-feedback loop while also providing a user interface for starting sessions and viewing summary statistics.

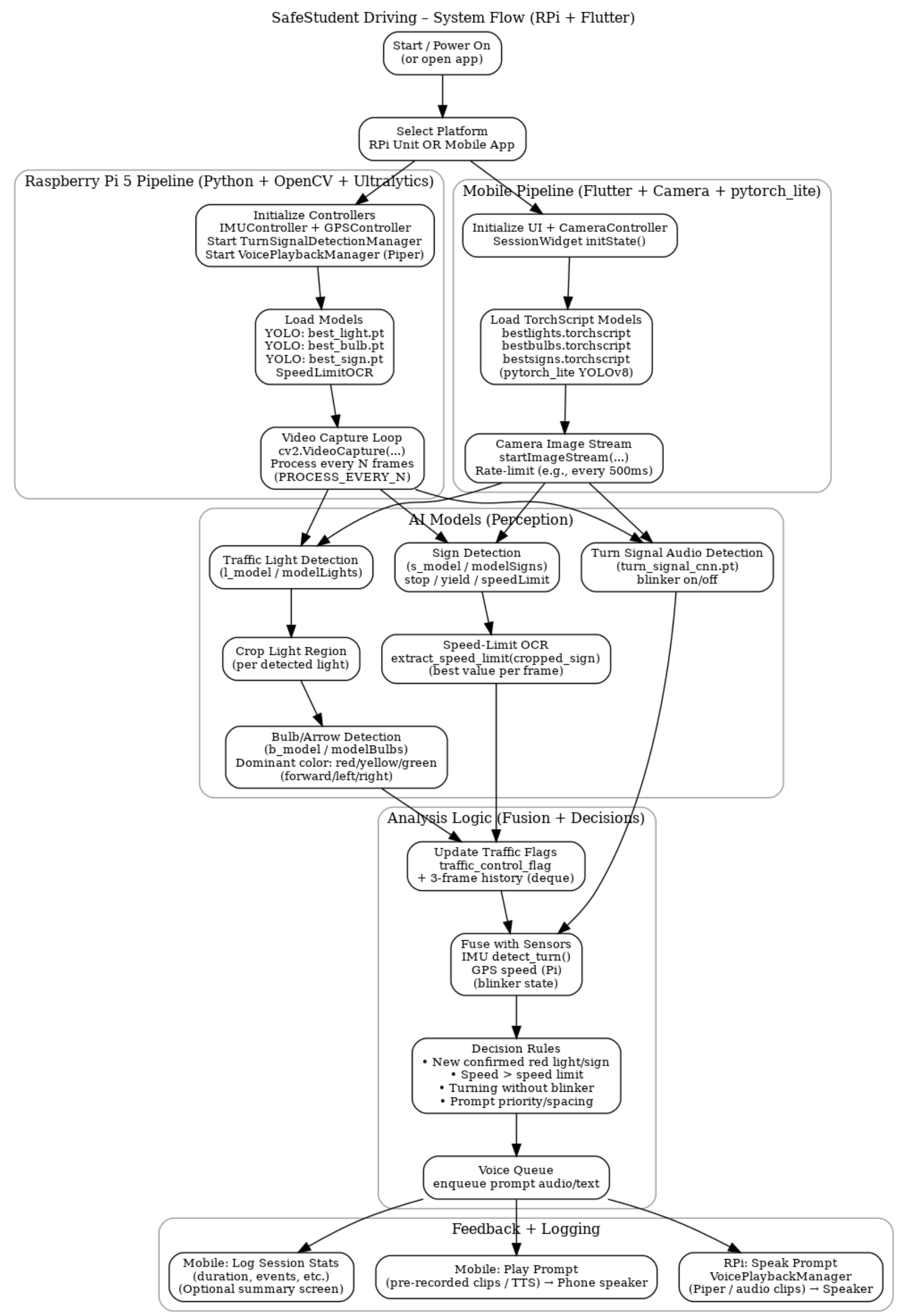

Figure 1. Overview of the solution

The AI models provide the system's perception. Three YOLO-based image models detect traffic lights, individual light bulbs, and road signs, while a separate CNN model classifies turn-signal audio [10]. Together, these models convert raw camera frames and microphone samples into structured events like "red light ahead" or "speed limit exceeded," which the logic layer can use.

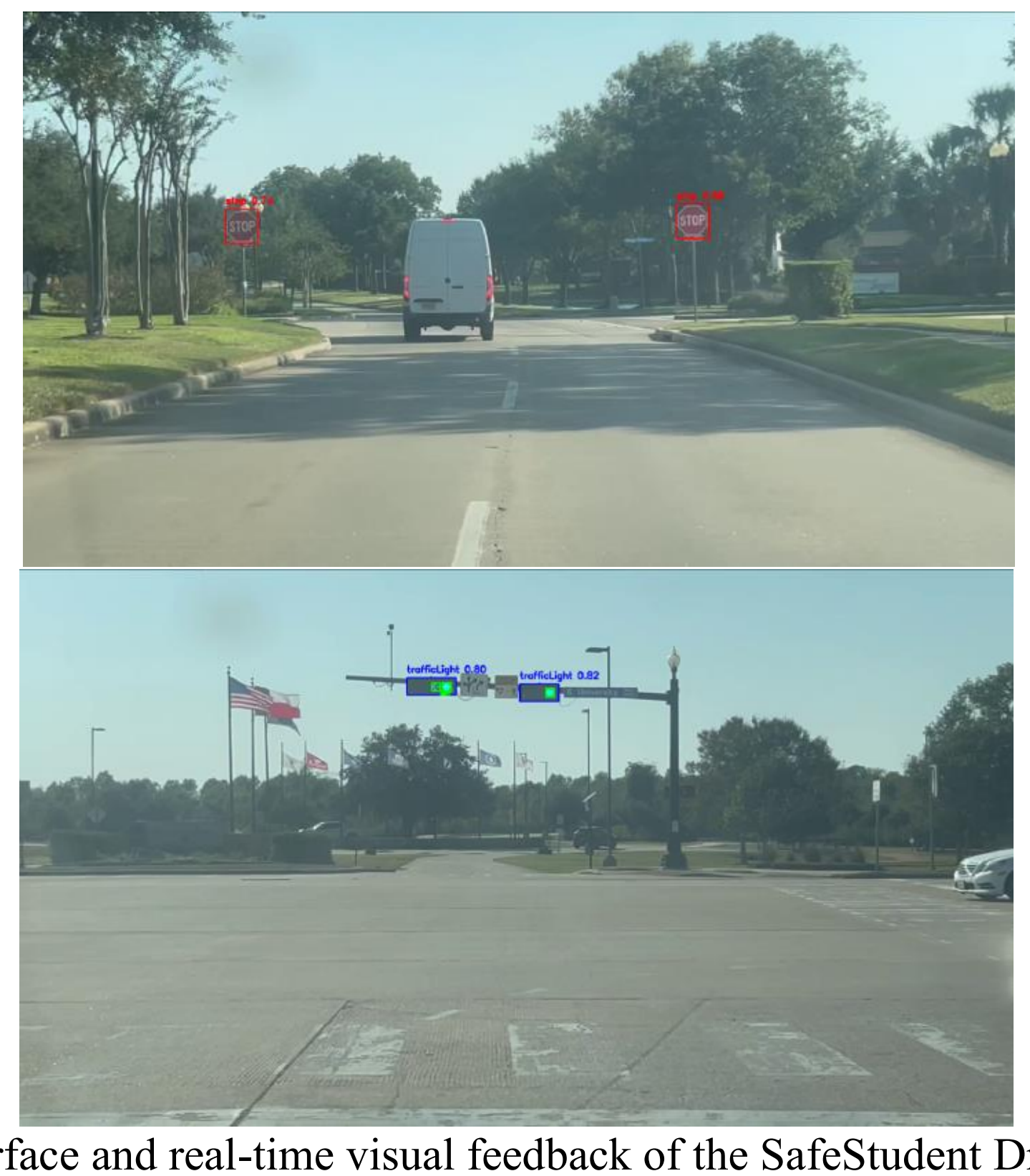

Figure 2. User interface and real-time visual feedback of the SafeStudent Driving system during operation

```
from ultralytics import YOLO
import cv2

from speed_limit_OCR import SpeedLimitOCR
from utils import crop_image
from light_utils import analyze_light_bulbs  # if you keep it in a separate file

# Load YOLO models
l_model = YOLO("models/image/best_light.pt")
b_model = YOLO("models/image/best_bulb.pt")
s_model = YOLO("models/image/best_sign.pt")

# OCR engine for speed-limit signs
speed_limit_ocr = SpeedLimitOCR(["en"])

# Thresholds
l_conf_threshold = 0.5
b_conf_threshold = 0.5
s_conf_threshold = 0.4
speed_limit_ocr_conf_threshold = 80
diff_conf = 0.1

# Flags are defined earlier in the file
# traffic_control_flag = {...}

cap = cv2.VideoCapture("tests/video/IMG_1409.mov")
count = 0
PROCESS_EVERY_N = 10

while cap.isOpened():
    ret, frame = cap.read()
    if not ret:
        break

    count += 1
    if count % PROCESS_EVERY_N != 0:
        continue

    current_traffic_control_flag = traffic_control_flag.copy()
    plot_box_list = []

    # ==========================
    # 1. Traffic light detection
```

```python
# ========================
l_results = l_model(frame, verbose=False)
l_class_names = l_model.names

for l_box in l_results[0].boxes:
    if l_box.conf[0] <= l_conf_threshold:
        continue

    l_box_name = l_class_names[int(l_box.cls[0].item())]
    plot_box_list.append((l_box, l_box_name, 0, 0))

    # Crop around the detected traffic light
    l_x, l_y, l_w, l_h = map(int, l_box.xywh[0])
    cropped_traffic_light_frame = crop_image(
        frame, (l_x - l_w / 2, l_y - l_h / 2, l_w, l_h)
    )

    # Detect bulbs inside the light
    b_results = b_model(cropped_traffic_light_frame, verbose=False)
    b_class_names = b_model.names

    dominant_color, dominant_color_box_list = analyze_light_bulbs(
        b_results, b_class_names, b_conf_threshold, diff_conf
    )

    if dominant_color.get("forward") == "red":
        current_traffic_control_flag["RedTrafficLight"] = True
    if dominant_color.get("left") == "red":
        current_traffic_control_flag["RedLeftTurnTrafficLight"] = True
    if dominant_color.get("right") == "red":
        current_traffic_control_flag["RedRightTurnTrafficLight"] = True

    offset_x = int(l_x - l_w / 2)
    offset_y = int(l_y - l_h / 2)
    plot_box_list.extend(
        (b_box, b_box_name, offset_x, offset_y)
        for (b_box, b_box_name) in dominant_color_box_list
    )

# ========================
# 2. Traffic sign detection + OCR
# ========================
s_results = s_model(frame, verbose=False)
s_class_names = s_model.names

best_speed_limit_conf = 0.0
best_speed_limit_value = None
```

```python
for s_box in s_results[0].boxes:
    if s_box.conf[0] <= s_conf_threshold:
        continue

    s_box_name = s_class_names[int(s_box.cls[0].item())]

    match s_box_name:
        case "stop":
            current_traffic_control_flag["StopTrafficSign"] = True
            plot_box_list.append((s_box, s_box_name, 0, 0))
        case "stopAhead":
            current_traffic_control_flag["StopAheadTrafficSign"] = True
            plot_box_list.append((s_box, s_box_name, 0, 0))
        case "yield":
            current_traffic_control_flag["YieldTrafficSign"] = True
            plot_box_list.append((s_box, s_box_name, 0, 0))
        case "yieldAhead":
            current_traffic_control_flag["YieldAheadTrafficSign"] = True
            plot_box_list.append((s_box, s_box_name, 0, 0))
        case _:
            # Treat as potential speed-limit sign
            s_x1, s_y1, s_x2, s_y2 = map(int, s_box.xyxy[0])
            cropped = frame[s_y1:s_y2, s_x1:s_x2]

            speed_limit_value = speed_limit_ocr.extract_speed_limit(
                cropped, speed_limit_ocr_conf_threshold
            )
            if speed_limit_value is not None:
                current_traffic_control_flag["SpeedLimitTrafficSign"] = True
                plot_box_list.append((s_box, s_box_name, 0, 0))

                s_box_conf = s_box.conf[0].item()
                if s_box_conf >= best_speed_limit_conf:
                    best_speed_limit_conf = s_box_conf
                    best_speed_limit_value = speed_limit_value
```

Figure 3. Core traffic-light and traffic-sign detection pipeline implemented in Python

In the shown code snippet, each processed frame first goes through traffic light detection with l_model. For every high-confidence light (l_box.conf[0] > l_conf_threshold), the program crops the region around the light and resizes it. That cropped image is then passed to b_model, which detects individual bulb shapes. The results (b_results, b_class_names) are fed to analyze_light_bulbs, which aggregates confidences per color and direction and returns a dominant_color dictionary plus a list of bulb boxes to visualize.

Next, the same frame is passed to s_model for sign detection. The loop branches on the predicted class name: stop and yield signs set the appropriate flags and are added to plot_box_list. Any other sign is treated as a potential speed-limit sign: the code crops the sign region, calls speed_limit_ocr.extract_speed_limit(...) with a confidence threshold, and records the best speed-limit value and corresponding detection confidence.

At the end of this stage, the system has a set of booleans for different traffic controls and, when available, a numeric speed limit.

The analysis logic functions as the system's central decision-making component. It continuously aggregates and interprets multi-modal inputs, including vision-based traffic control detections, GPS-derived vehicle speed, IMU-based turning information, and turn-signal audio cues [11]. To improve robustness, the system applies temporal smoothing to noisy visual detections and maintains an internal state to track which traffic controls are currently active.

By fusing perception results with motion and signal data, the analysis module determines appropriate moments to enqueue voice prompts for the driver. A history-based confirmation mechanism and internal flags are used to suppress redundant, contradictory, or premature alerts, ensuring that guidance is issued only when sufficient confidence is established [12]. During inference, system behavior is validated through both console-level diagnostic outputs and visual driving scene representations, including stock driving images annotated with subtitles corresponding to detected events and generated prompts. This design enables reliable, context-aware driver assistance while minimizing alert fatigue.

```python
from collections import deque

traffic_control_flag = {
    "RedTrafficLight": False,
    "RedLeftTurnTrafficLight": False,
    "RedRightTurnTrafficLight": False,
    "StopTrafficSign": False,
    "StopAheadTrafficSign": False,
    "YieldTrafficSign": False,
    "YieldAheadTrafficSign": False,
    "SpeedLimitTrafficSign": False,
}


traffic_control_flag_history = deque(maxlen=3)
for _ in range(3):
    traffic_control_flag_history.append(traffic_control_flag.copy())

last_confirmed_traffic_control_flag = traffic_control_flag.copy()

# ... inside the main loop, after current_traffic_control_flag is populated ...

traffic_control_flag_history.append(current_traffic_control_flag)

# ---- Turn-signal coaching ----
if (
    not turn_signal_detection_manager.is_blinker_on
    and imu_controller.detect_turn() != "Straight"
):
    voice_playback_manager.speak_queued(
        "mobile_app/max_driving/assets/audios/turn_signal.wav"
    )

# ---- Speeding coaching ----
if (
    current_traffic_control_flag["SpeedLimitTrafficSign"]
    and not last_confirmed_traffic_control_flag["SpeedLimitTrafficSign"]
):
    gps_controller.update()
    driving_mph = gps_controller.get_speed_mph()
    if best_speed_limit_value is not None and driving_mph > best_speed_limit_value:
        voice_playback_manager.speak_queued(
            "mobile_app/max_driving/assets/audios/speed_limit.wav"
        )
    last_confirmed_traffic_control_flag["SpeedLimitTrafficSign"] = (
        current_traffic_control_flag["SpeedLimitTrafficSign"]
    )
```

```python
# ---- Traffic-control coaching with 3-frame confirmation ----
for key in current_traffic_control_flag:
    # Detected 3 frames in a row → turn "on"
    if (
        all(history[key] for history in traffic_control_flag_history)
        and not last_confirmed_traffic_control_flag[key]
    ):
        match key:
            case "RedTrafficLight":
                voice_playback_manager.speak_queued(
                    "mobile_app/max_driving/assets/audios/red_light.wav"
                )
            case "RedLeftTurnTrafficLight":
                voice_playback_manager.speak_queued(
                    "mobile_app/max_driving/assets/audios/red_left_light.wav"
                )
            .........

        last_confirmed_traffic_control_flag[key] =
current_traffic_control_flag[key]

    # Not detected 3 frames in a row → turn "off" (for selected keys)
    if (
        all(not history[key] for history in traffic_control_flag_history)
        and last_confirmed_traffic_control_flag[key]
    ):
        match key:
            case "RedTrafficLight":
                voice_playback_manager.speak_queued(
                    "mobile_app/max_driving/assets/audios/red_light_off.wav"
                )
            case "RedLeftTurnTrafficLight":
                voice_playback_manager.speak_queued(
                    "mobile_app/max_driving/assets/audios/red_left_light_off.wav"
                )

        last_confirmed_traffic_control_flag[key] =
current_traffic_control_flag[key]
```

Figure 4. Temporal confirmation and alert-triggering logic for traffic-control detection

The analysis logic begins by initializing traffic_control_flag with keys for each type of traffic control and storing three copies in traffic_control_flag_history. For every processed frame, the system creates a fresh current_traffic_control_flag (starting as a copy of the default) and sets entries to True when detections occur (e.g., StopTrafficSign, RedTrafficLight, SpeedLimitTrafficSign). This per-frame snapshot is appended to the deque.

To avoid reacting to single-frame glitches, the code looks for three consecutive frames all reporting the same condition. If, for a given key, all three entries in traffic_control_flag_history are True and that condition was previously False in last_confirmed_traffic_control_flag, the code considers it a confirmed event and plays the matching audio (for example, red_light.wav or stop.wav). Conversely, if a condition disappears for three frames in a row, the system may play an “off” sound (e.g., red_light_off.wav) and update last_confirmed_traffic_control_flag accordingly.

Additional checks use imu_controller.detect_turn() and turn_signal_detection_manager.is_blinker_on to detect turns without signals, triggering a dedicated prompt for that case, and GPS speed is compared against best_speed_limit_value to detect speeding.

The hardware component connects the software to real-world devices. On the Raspberry Pi, the system runs Python code directly on the board with a USB camera, microphone, GPS, IMU, and speaker [13]. On mobile, a Flutter app uses the phone’s built-in sensors and camera to implement a similar perception and feedback loop.

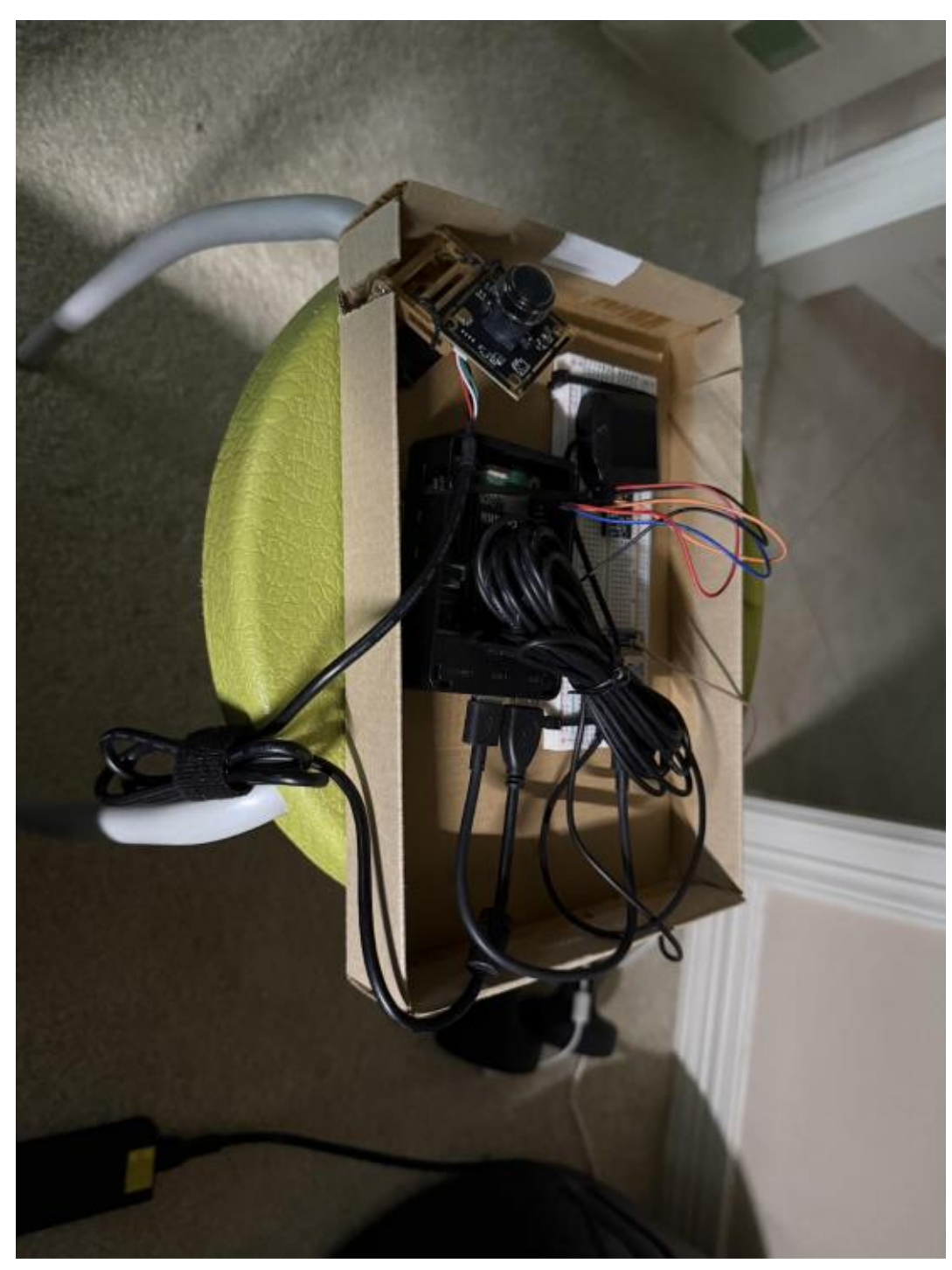

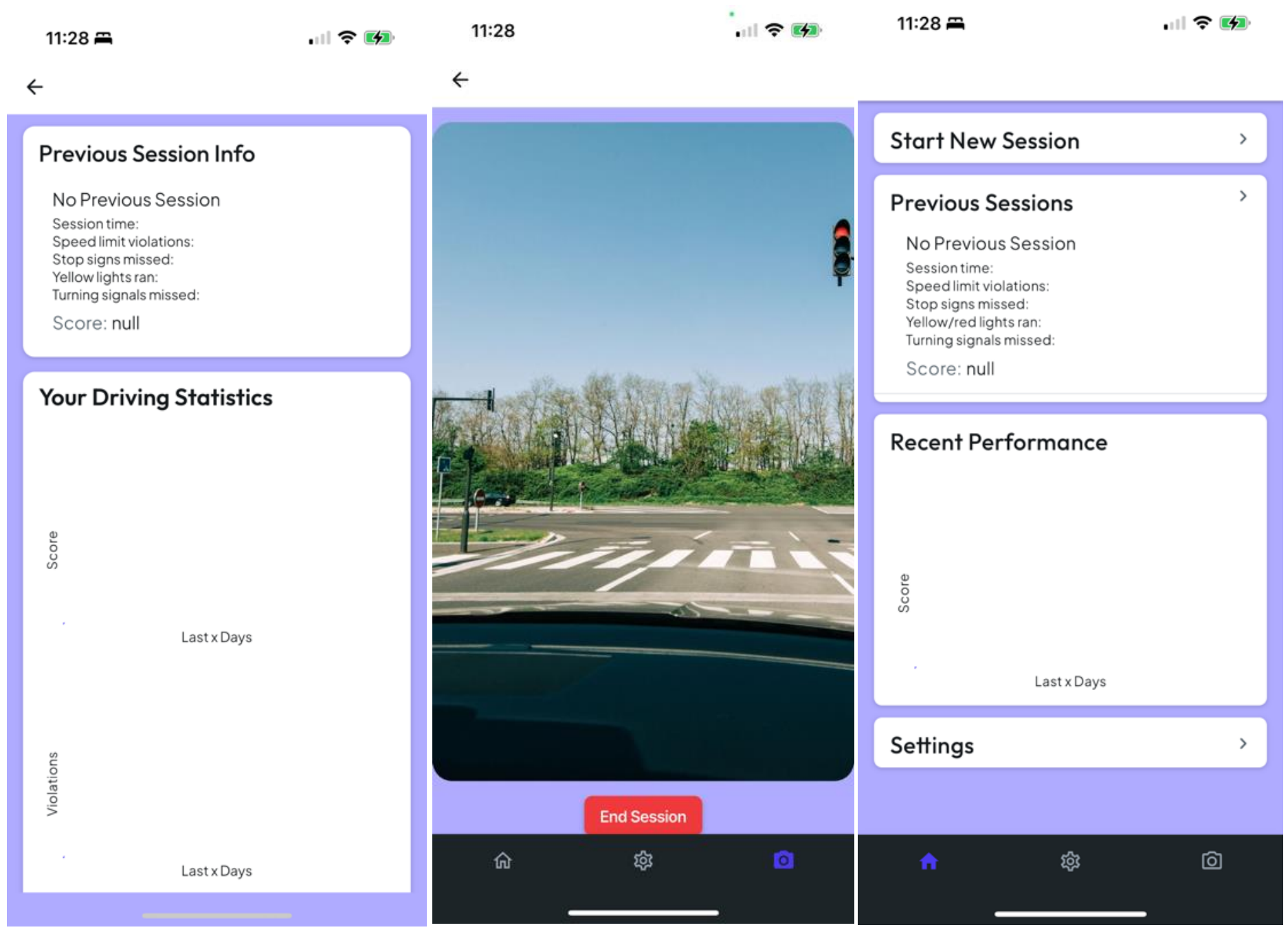

Figure 5. Mobile application interface showing live camera preview and system status indicators

```dart
import 'dart:io';
import 'package:camera/camera.dart';
import 'package:flutter/material.dart';
import 'package:pytorch_lite/pytorch_lite.dart';

// ---- Model loading and frame processing ----

ModelObjectDetection? modelBulbs;
ModelObjectDetection? modelLights;
ModelObjectDetection? modelSigns;

Future<void> loadModels() async {
  try {
    modelBulbs = await PytorchLite.loadObjectDetectionModel(
      'assets/models/bestbulbs.torchscript',
      6,
      640,
      640,
      labelPath: 'assets/labels/labels_bulbs.txt',
      objectDetectionModelType: ObjectDetectionModelType.yolov8,
    );

    modelLights = await PytorchLite.loadObjectDetectionModel(
      'assets/models/bestlights.torchscript',
      7,
      640,
      640,
      labelPath: 'assets/labels/labels_lights.txt',
      objectDetectionModelType: ObjectDetectionModelType.yolov8,
    );

    modelSigns = await PytorchLite.loadObjectDetectionModel(
      'assets/models/bestsigns.torchscript',
      18,
      640,
      640,
      labelPath: 'assets/labels/labels_signs.txt',
      objectDetectionModelType: ObjectDetectionModelType.yolov8,
    );

    print('Models loaded successfully');
  } catch (e) {
    print('Error loading models: $e');
  }
}

Future<void> processFrame(CameraImage image, int rotation) async {
```

```dart
  if (modelBulbs == null || modelLights == null || modelSigns == null) {
    await loadModels();
  }

  // Bulbs
  try {
    final bulbs = await modelBulbs?.getCameraImagePrediction(
      image,
      rotation: rotation,
      minimumScore: 0.5,
      iOUThreshold: 0.3,
    );
    print('Detected bulbs: ${bulbs?.length ?? 0}');
  } catch (e) {
    print('Error predicting bulbs: $e');
  }

  // Lights
  try {
    final lights = await modelLights?.getCameraImagePrediction(
      image,
      rotation: rotation,
      minimumScore: 0.5,
      iOUThreshold: 0.3,
    );
    print('Detected lights: ${lights?.length ?? 0}');
  } catch (e) {
    print('Error predicting lights: $e');
  }

  // Signs
  try {
    final signs = await modelSigns?.getCameraImagePrediction(
      image,
      rotation: rotation,
      minimumScore: 0.4,
      iOUThreshold: 0.3,
    );
    print('Detected signs: ${signs?.length ?? 0}');
  } catch (e) {
    print('Error predicting signs: $e');
  }
}

// ---- Session screen with camera integration ----

class SessionWidget extends StatefulWidget {
```

```dart
  const SessionWidget({super.key});

  @override
  State<SessionWidget> createState() => _SessionWidgetState();
}

class _SessionWidgetState extends State<SessionWidget> {
  late CameraController _cameraController;
  late Future<void> _initializeControllerFuture;
  DateTime? lastFrameTimestamp;

  @override
  void initState() {
    super.initState();

    availableCameras().then((cameras) {
      _cameraController = CameraController(
        cameras.first,
        ResolutionPreset.medium,
      );
      _initializeControllerFuture = _cameraController.initialize();
      setState(() {});
    });
  }

  @override
  void dispose() {
    _cameraController.dispose();
    super.dispose();
  }

  @override
  Widget build(BuildContext context) {
    return FutureBuilder<void>(
      future: _initializeControllerFuture,
      builder: (context, snapshot) {
        if (snapshot.connectionState != ConnectionState.done) {
          return const Center(child: CircularProgressIndicator());
        }

        Widget cameraPreview = CameraPreview(_cameraController);
        int rotation = 0;

        if (Platform.isAndroid) {
          cameraPreview = RotatedBox(
            quarterTurns: 1,
            child: cameraPreview,
```

```
          );
          rotation = 90;
        }

        // Start camera stream and send frames for processing
        _cameraController.startImageStream((CameraImage image) {
          if (lastFrameTimestamp != null &&
              DateTime.now().difference(lastFrameTimestamp!) <
                  const Duration(milliseconds: 500)) {
            return;
          }

          processFrame(image, rotation);
          lastFrameTimestamp = DateTime.now();
        });

        return AspectRatio(
          aspectRatio: _cameraController.value.aspectRatio,
          child: ClipRRect(
            borderRadius: BorderRadius.circular(20.0),
            child: cameraPreview, // real preview; swap for a placeholder if needed
          ),
        );
      },
    );
  }
}
```

Figure 6. Flutter-based real-time camera stream processing and model inference workflow

In the Flutter component, the hardware integration begins in initState(), where the app calls availableCameras() to get a list of cameras on the device and then creates a CameraController with ResolutionPreset.medium. The _initializeControllerFuture is used by a FutureBuilder to wait until the camera is fully initialized before showing anything on screen. Once ready, the widget builds a CameraPreview wrapped in a rounded ClipRRect so it visually matches the “session” card. On Android, the preview is rotated by 90 degrees using RotatedBox to correct the camera orientation, and the same rotation value is passed into processFrame() so the YOLO models interpret the image correctly [14].

The key line for live processing is _cameraController.startImageStream(...), which delivers a continuous stream of CameraImage frames. To avoid overloading the phone, the code rate-limits processing by checking lastFrameTimestamp and only calling processFrame(image, rotation) every 500 milliseconds. Inside processFrame, the three TorchScript YOLO models are lazily loaded if needed and then run on the current frame to detect bulbs, lights, and signs, mirroring the perception behavior of the Raspberry Pi version.

## 4. EXPERIMENT

### 4.1. Experiment 1

We needed to test how accurately the system detects and reads speed-limit and regulatory signs in real driving footage, especially under different lighting and distances, because incorrect sign readings could lead to unsafe feedback.

To evaluate sign and speed-limit performance, we will create a labeled dataset of road clips recorded from the Raspberry Pi and phone, covering daytime, sunset, and nighttime. For each clip, we will annotate the ground-truth signs and speed-limit values by hand (e.g., “Stop ahead,” “Yield,” “Speed limit exceeded”). Then we will run the full sign pipeline from main()-YOLO sign detection followed by SpeedLimitOCR.extract_speed_limit(...)-on every frame sampled at fixed intervals (for example, every 10th frame). For each annotated sign, we will measure whether the system (a) detected it and (b) correctly classified it and, for speed-limit signs, whether it extracted the right numeric value.

| Clip | Lighting | End-to-end Correctness (%) |
|---|---|---|
| D1 | Day | 92 |
| D2 | Day | 90 |
| D3 | Day | 93 |
| D4 | Day | 88 |
| S1 | Sunset | 86 |
| S2 | Sunset | 84 |
| S3 | Sunset | 82 |
| S4 | Sunset | 85 |
| N1 | Night | 74 |
| N2 | Night | 78 |
| N3 | Night | 71 |
| N4 | Night | 76 |

Figure 7. Experimental setup and evaluation results for speed-limit and regulatory sign detection under different lighting conditions

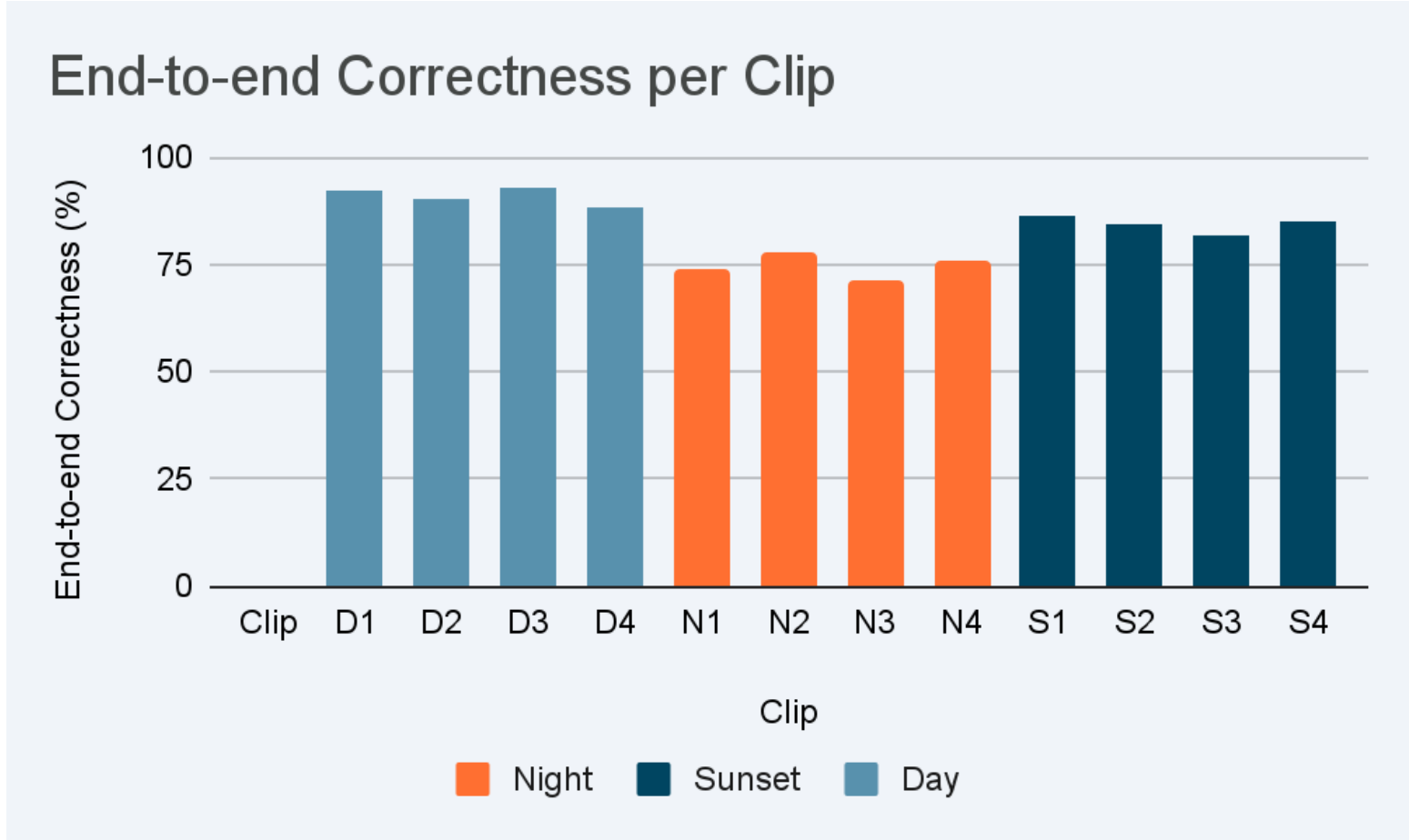

Figure 8. Quantitative performance comparison of sign-detection accuracy across lighting conditions

Across 12 real-driving clips, the system’s end-to-end sign pipeline correctness averaged 83.25% with a median of 84.5%. The lowest clip score was 71% (Night clip N3), and the highest was 93% (Day clip D3). Performance was strongest in daylight (mean 90.75%) and declined at sunset (mean 84.25%) and nighttime (mean 74.75%). The biggest surprise was how sharply nighttime dropped compared to sunset: YOLO often still detected a sign, but the OCR numeric read failed more frequently due to glare from headlights, reflective sign material, motion blur, and lower effective resolution when signs were farther away. Because the pipeline requires both detection + correct classification + correct OCR to count as correct, OCR errors disproportionately reduce the final score. In practice, lighting and viewing distance had the largest effect on results, especially for speed-limit signs, where a single misread digit makes the output wrong.

### 4.2. Experiment 2

We also wanted to test how reliably the turn-signal audio model detects real blinker clicks in different cars and noise environments, since missed or false detections could lead to incorrect coaching about turn-signal use.

To evaluate the turn-signal detector, we will record a dataset of driving segments using the same microphone setup as in the Raspberry Pi system. Each segment will be labeled with ground-truth events: "signal on," "signal off," plus background conditions (windows open/closed, music on/off, highway vs. city). The TurnSignalDetectionManager (initialized in main() with turn_signal_cnn.pt) will run continuously on these recordings, producing a binary prediction ("blinker on" vs. "off") over time. We will align predictions with the ground-truth timeline and compute metrics such as true positive rate, false positive rate, and F1-score, both overall and under each noise condition.

| Car | Silence | Music | Conversation | Music + Conversation |
|---|---|---|---|---|
| Car A | 0.95 | 0.90 | 0.88 | 0.82 |
| Car B | 0.93 | 0.87 | 0.85 | 0.80 |
| Car C | 0.91 | 0.86 | 0.84 | 0.78 |

Figure 9. Figure of experiment 2

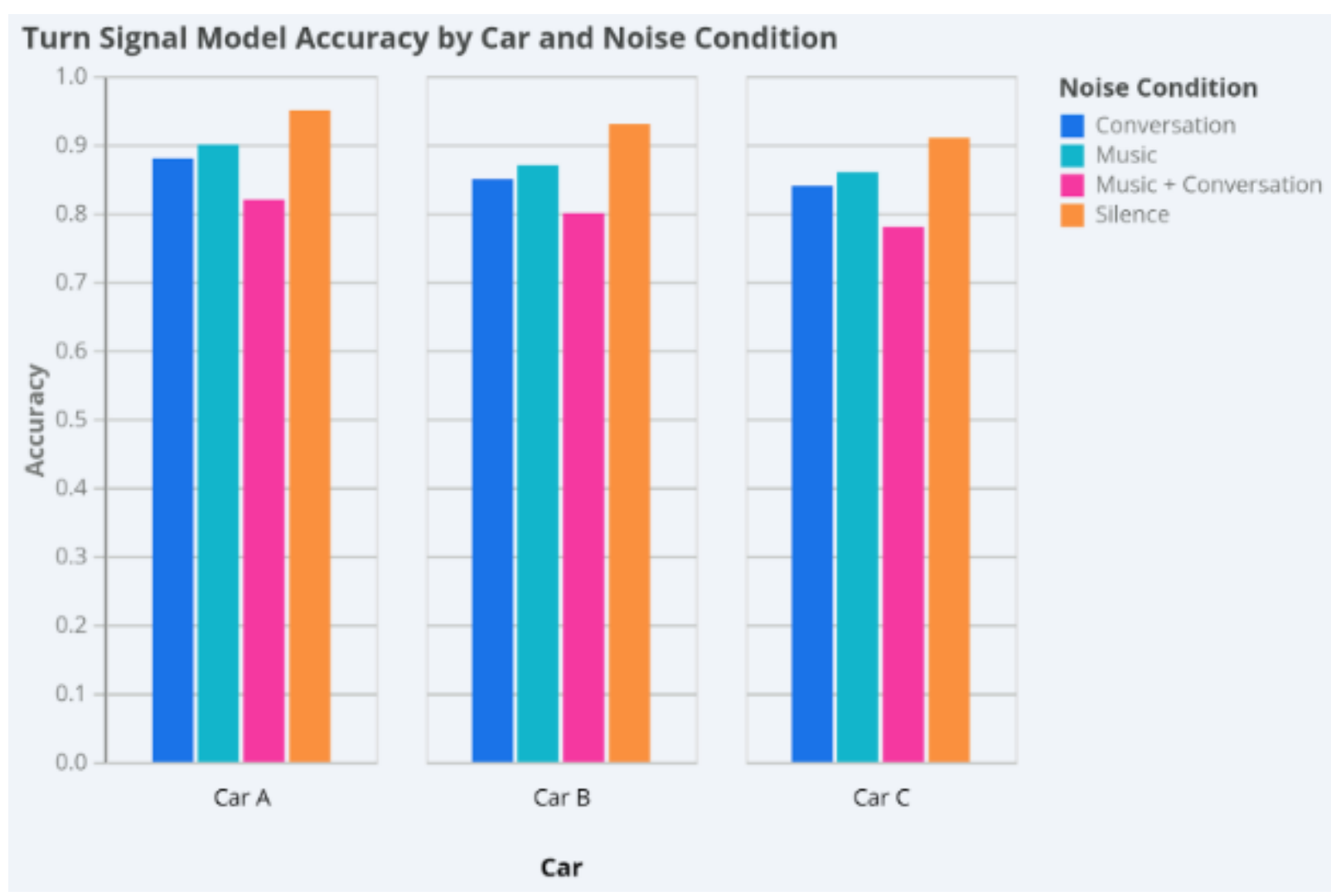

Figure 10. Figure of experiment

Across all 12 recordings (3 cars × 4 conditions), the turn-signal detector achieved a mean F1-score of 0.866 with a median of 0.865. The lowest result was 0.78 (Car C with music + conversation), while the highest was 0.95 (Car A in silence). As expected, performance was best in quiet cabins (mean 0.93) and degraded as background audio increased (music mean 0.877, conversation mean 0.857, music+conversation mean 0.80). The main pattern is that false positives rise when rhythmic audio (music) overlaps the blinker's click cadence, and false negatives rise when clicks are partially masked by speech or when the microphone gain is tuned low to avoid clipping. Differences across cars also matter because blinker "click" timbre varies by vehicle, so a model trained on one set of clicks may generalize imperfectly. Overall, ambient noise level was the biggest driver of F1 changes, followed by vehicle-to-vehicle sound variation, suggesting the largest improvement would come from training with more diverse cars and more noisy examples.

## 5. RELATED WORK

Ferreira Júnior et al. (2017) propose a smartphone-based driver behavior profiling system that uses built-in sensors (accelerometer, gyroscope, GPS) plus machine-learning classifiers to categorize driving style as aggressive, normal, or calm [4]. The approach is effective for low-cost monitoring and achieves good classification performance over different sensor combinations (Ferreira Júnior et al., 2017). However, it mainly focuses on overall style rather than specific road elements like stoplights or speed-limit signs, and it does not provide real-time, context-aware coaching to the driver. SafeStudent Driving improves this by adding camera-based traffic control detection, turn-signal audio monitoring, and immediate spoken feedback aimed at teaching novice drivers safer habits.

Brahim et al. (2022) present a smartphone-based sensing framework that uses machine learning to classify driving behavior by combining vehicle-dynamics and driver-dynamics signals, such as acceleration patterns and steering motions [5]. Their pipeline focuses on building accurate classifiers and discusses how different sensor choices affect performance (Brahim et al., 2022). While powerful for behavior recognition, the system largely treats the road environment as a black box: it does not explicitly recognize traffic lights, stop signs, or post speed limits, and it is not targeted specifically at teen drivers. SafeStudent Driving extends this idea by fusing sensor-based behavior cues with explicit visual understanding of traffic controls and by delivering simple, pedagogical audio prompts in real time.

Ramachandra et al. (2022) implement an advanced driver assistance system (ADAS) on a Raspberry Pi that performs lane detection, blind-spot monitoring, forward-collision warning, and pedestrian detection using camera input and image processing [6]. Their hardware prototype shows that low-cost, Pi-based ADAS features are feasible and can enhance driver awareness (Ramachandra et al., 2022). However, the system focuses on collision avoidance rather than on teaching long-term driving habits, and it does not explicitly track compliance with traffic controls (such as speed-limit signs or turn-signal usage). SafeStudent Driving builds on this Raspberry-Pi ADAS direction but targets novice drivers, adds speed-limit OCR and turn-signal audio detection, and emphasizes coaching via repeated, gentle voice prompts rather than only warning about imminent hazards.

## 6. CONCLUSIONS

Although SafeStudent Driving demonstrates that low-cost hardware and mobile devices can provide helpful coaching for novice drivers, several limitations remain. First, detection accuracy still depends strongly on lighting, viewing distance, weather, and camera quality. Signs at night, in rain, or far from the lens are more likely to be missed or misread and turn-signal audio detection can be confused by loud music or high cabin noise. With additional development time, future work would expand the training and test datasets to include more nighttime footage, adverse weather, and varied vehicle interiors, followed by targeted fine-tuning and threshold calibration for these conditions. Second, the current feedback audio relies on relatively simple text-to-speech or pre-recorded clips, which can sound repetitive and reduce user engagement over time. Future iterations could incorporate higher-quality neural TTS and more varied phrasing while preserving brevity [16]. Finally, the Raspberry Pi prototype is still physically bulky; a more product-like design would integrate the components into a smaller enclosure and explore more power-efficient hardware or dedicated accelerators to reduce size, heat, and power draw.

SafeStudent Driving shows that a combination of computer vision, audio analysis, and simple sensor fusion can turn everyday devices into effective coaching tools for new drivers. By focusing on clear, real-time feedback about traffic controls and turn-signal use, the system aims to help teens develop safer habits from their earliest miles on the road.